\documentclass[]{spie}  %>>> use for US letter paper
\pdfoutput=1
\usepackage{fixltx2e}
\usepackage[T1]{fontenc}
\usepackage[ascii]{inputenc}

\usepackage[]{graphicx}
\usepackage[]{hyperref}

\usepackage{aas_macros}

\usepackage[noblocks]{authblk}
\makeatletter
\renewcommand\AB@authnote[1]{\textsuperscript{\itshape #1}}
\renewcommand\AB@affilnote[1]{\textsuperscript{\itshape #1}}
\renewcommand\AB@affilsepx{;\protect\\\protect\Affilfont}
\makeatother
\usepackage{etoolbox}
\usepackage{alphalph}
\newcounter{affils}
\newcommand\affillist\relax
\AtEndPreamble{\affillist}
\newcommand\affilcounter[2]{
        \refstepcounter{affils}
        \newcounter{#1}
        \setcounter{#1}{\value{affils}}
        \expandafter\renewcommand\csname the#1\endcsname{\alphalph{\value{#1}}}
        \appto\affillist{\affil[\csuse{the#1}]{#2}}%
        \csgdef{#1}{\csuse{the#1}}
}

\title{BLASTbus electronics: general-purpose readout and control for balloon-borne experiments} 

\affilcounter{uoftp}{Department of Physics, University of Toronto, Toronto, ON, Canada}
\affilcounter{cardiff}{School of Physics and Astronomy, Cardiff University, Cardiff, UK}
\affilcounter{ubc}{Department of Physics \& Astronomy, University of British Columbia, Vancouver, BC, Canada}
\affilcounter{penn}{Department of Physics \& Astronomy, University of Pennsylvania, Philadelphia, PA, USA}
\affilcounter{cit}{Division of Physics, Mathematics \& Astronomy, California Institute of Technology, Pasadena, CA, USA}
\affilcounter{jpl}{Jet Propulsion Laboratory, Pasadena, CA, USA}
\affilcounter{cita}{Canadian Institute for Theoretical Astrophysics, Toronto, ON, Canada}
\affilcounter{cifar}{Canadian Institute for Advanced Research, Toronto, ON, Canada}
\affilcounter{cwru}{Department of Physics, Case Western Reserve University, Cleveland, OH, USA}
\affilcounter{ukzn}{Astrophysics \& Cosmology Research Unit, University of KwaZulu-Natal, Durban, South Africa}
\affilcounter{icl}{Theoretical Physics, Blackett Laboratory, Imperial College, London, UK}
\affilcounter{uoft}{Department of Astronomy \& Astrophysics, University of Toronto, Toronto, ON, Canada}
\affilcounter{ciera}{Center for Interdisciplinary Exploration and Research in Astrophysics, Northwestern University, Evanston, IL, USA}
\affilcounter{pton}{Department of Physics, Princeton University, Princeton, NJ, USA}
\affilcounter{nagoya}{Institute for Advanced Research, Nagoya University, Furo-cho, Chikusa-ku, Nagoya-shi, Japan}
\affilcounter{ptona}{Department of Astrophysical Sciences, Princeton University, Princeton, NJ, USA}
\affilcounter{nist}{National Institute of Standards and Technology, Boulder, CO, USA}
\affilcounter{stanford}{Department of Physics, Stanford University, Stanford, CA, USA}
\affilcounter{kipacslac}{Kavli Institute for Particle Astrophysics and Cosmology, SLAC National Accelerator Laboratory, Menlo Park, CA, USA}
\affilcounter{brown}{Department of Physics, Brown University, Providence, RI, USA}
\affilcounter{kicp}{Kavli Institute for Cosmology, University of Cambridge, Cambridge, UK}
\affilcounter{nw}{Department of Physics \& Astronomy, Northwestern University, Evanston, IL, USA}
\affilcounter{fred}{Instituto de Astrof\'isica de Canarias, La Laguna, Tenerife, Spain}
\affilcounter{fredd}{Departamento de Astrof\'isica, Universidad de La Laguna, La Laguna, Tenerife, Spain}
\affilcounter{ucl}{Department of Physics and Astronomy, University College London, London, UK}
\affilcounter{juan}{Institut d'Astrophysique Spatiale, CNRS \& Universit\'e Paris-Sud, Orsay, France}
\affilcounter{miami}{Department of Physics, University of Miami, Coral Gables, FL, USA}
\affilcounter{derek}{Jeremiah Horrocks Institute of Maths, Physics and Astronomy, University of Central Lancashire, Preston, UK}

\author[\uoftp]{S.\,J.~Benton}
\authorinfo{S.J.~Benton: E-mail: sbenton@physics.utoronto.ca, Telephone: 1 416 946 0946\\
Copyright 2014 Society of Photo-Optical Instrumentation Engineers. One print or electronic copy may be made for personal use only. Systematic reproduction and distribution, duplication of any material in this paper for a fee or for commercial purposes, or modification of the content of the paper are prohibited.
}
\author[\cardiff]{P.\,A.~Ade}
\author[\ubc]{M.~Amiri}
\author[\penn]{F.\,E.~Angil\`e}
\author[\cit,\jpl]{J.\,J.~Bock}
\author[\cita,\cifar]{J.\,R.~Bond}
\author[\cwru]{S.\,A.~Bryan}
\author[\ukzn]{H.\,C.~Chiang}
\author[\icl]{C.\,R.~Contaldi}
\author[\cit,\jpl]{B.\,P.~Crill}
\author[\penn]{M.\,J.~Devlin}
\author[\penn]{B.~Dober}
\author[\cit,\jpl]{O.\,P.~Dor\'e}
\author[\jpl]{C.\,D.~Dowell}
\author[\cita,\uoft]{M.~Farhang}
\author[\cit]{J.\,P.~Filippini}
\author[\ciera,\uoft]{L.\,M.~Fissel}
\author[\pton]{A.\,A.~Fraisse}
\author[\nagoya]{Y.~Fukui}
\author[\penn]{N.~Galitzki}
\author[\pton]{A.\,E.~Gambrel}
\author[\uoft]{N.\,N.~Gandilo}
\author[\cit]{S.\,R.~Golwala}
\author[\pton]{J.\,E.~Gudmundsson}
\author[\ubc,\cifar]{M.~Halpern}
\author[\ptona]{M.~Hasselfield}
\author[\nist]{G.\,C.~Hilton}
\author[\jpl]{W.\,A.~Holmes}
\author[\cit]{V.\,V.~Hristov}
\author[\stanford,\kipacslac,\nist]{K.\,D.~Irwin}
\author[\pton]{W.\,C.~Jones}
\author[\pton]{Z.\,D.~Kermish}
\author[\penn]{J.~Klein}
\author[\brown]{A.\,L.~Korotkov}
\author[\stanford,\kipacslac]{C.\,L.~Kuo}
\author[\kicp]{C.\,J.~MacTavish}
\author[\cit]{P.\,V.~Mason}
\author[\nw]{T.\,G.~Matthews}
\author[\jpl]{K.\,G.~Megerian}
\author[\cit]{L.~Moncelsi}
\author[\cit]{T.\,A.~Morford}
\author[\cit]{T.\,K.~Mroczkowski}
\author[\cwru]{J.\,M.~Nagy}
\author[\uoftp,\cifar,\uoft]{C.\,B.~Netterfield}
\author[\ciera]{G.~Novak}
\author[\cardiff]{D.~Nutter}
\author[\cit,\jpl]{R.~O'Brient}
\author[\kipacslac]{R.\,W.~Ogburn~IV}
\author[\cardiff]{E.~Pascale}
\author[\fred,\fredd]{F.~Poidevin}
\author[\pton]{A.\,S.~Rahlin}
\author[\nist]{C.\,D.~Reintsema}
\author[\cwru]{J.\,E.~Ruhl}
\author[\jpl]{M.\,C.~Runyan}
\author[\ucl]{G.~Savini}
\author[\ubc]{D.~Scott}
\author[\uoft]{J.\,A.~Shariff}
\author[\juan,\uoft]{J.\,D.~Soler}
\author[\miami]{N.\,E.~Thomas}
\author[\jpl]{A.~Trangsrud}
\author[\penn]{M.\,D.~Truch}
\author[\cardiff]{C.\,E.~Tucker}
\author[\brown]{G.\,S.~Tucker}
\author[\cit]{R.\,S.~Tucker}
\author[\jpl]{A.\,D.~Turner}
\author[\derek]{D.~Ward-Thompson}
\author[\jpl]{A.\,C.~Weber}
\author[\ubc]{D.\,V.~Wiebe}
\author[\pton]{E.\,Y.~Young}

\begin{document} 
\maketitle 

%%%%%%%%%%%%%%%%%%%%%%%%%%%%%%%%%%%%%%%%%%%%%%%%%%%%%%%%%%%%% 
\begin{abstract}
We present the second generation BLASTbus electronics. The primary purposes of this system are detector readout, attitude control, and cryogenic housekeeping, for balloon-borne telescopes. Readout of neutron transmutation doped germanium (NTD-Ge) bolometers requires low noise and parallel acquisition of hundreds of analog signals. Controlling a telescope's attitude requires the capability to interface to a wide variety of sensors and motors, and to use them together in a fast, closed loop. To achieve these different goals, the BLASTbus system employs a flexible motherboard-daughterboard architecture. The programmable motherboard features a digital signal processor (DSP) and field-programmable gate array (FPGA), as well as slots for three daughterboards. The daughterboards provide the interface to the outside world, with versions for analog to digital conversion, and optoisolated digital input/output. With the versatility afforded by this design, the BLASTbus also finds uses in cryogenic, thermometry, and power systems. For accurate timing control to tie everything together, the system operates in a fully synchronous manner. BLASTbus electronics have been successfully deployed to the South Pole, and flown on stratospheric balloons.
\end{abstract}

%>>>> Include a list of keywords after the abstract 


%%%%%%%%%%%%%%%%%%%%%%%%%%%%%%%%%%%%%%%%%%%%%%%%%%%%%%%%%%%%%
\section{Introduction}
\label{sec:intro}
Balloon-borne telescopes have many different sensors to read and subsystems to control. The BLASTbus system was developed to accomplish both detector readout and telescope pointing for the BLAST\cite{pascale2008} experiment. The system also found a niche in cryogenic control and housekeeping. For the BLASTPol\cite{pascale2012} and \textsc{Spider}\cite{rahlin2014} experiments, a second-generation system was designed to lower power consumption while increasing modularity and usability. 

In addition to BLAST, BLASTpol, and \textsc{Spider}, the BLASTbus electronics have been used by several other experiments. The first-generation system flew on the EBEX\cite{oxley2004} telescope for attitude control. For cryogenic housekeeping in ground-based telescopes, the first-generation system has been deployed by ACT\cite{swetz2011} and BICEP2\cite{ogburn2010}. The second-generation has been deployed by the Keck Array\cite{staniszewski2012} and will soon be deployed by BICEP3\cite{kuo2013}.

This work outlines the architecture of the second-generation BLASTbus electronics (Sec.~\ref{sec:arch}), then examines its three main applications: bolometer readout (Sec.~\ref{sec:ntd_readout}); attitude control (Sec.~\ref{sec:attitude}); and cryogenic housekeeping (Sec.~\ref{sec:housekeeping}).

\section{BLASTbus System Architecture}
\label{sec:arch}

\subsection{The BLASTbus}
\label{sec:bbus}
The BLASTbus is a half-duplex RS-485\footnote{TIA/EIA-485-A} serial bus. The master node of the bus is a PCI board (Sec.~\ref{sec:bbcpci}) controlled by a computer, and the slave nodes are BLASTbus motherboards (Sec.~\ref{sec:motherboard}). An always-driven clock at up to 5\,MHz is used for serial communications and to synchronize the motherboards. Nominally, a 4\,MHz clock is used, but for MCE synchronization (Sec.~\ref{sec:sync}) the MCE's 5\,MHz clock can be used instead. The bus is optoisolated at both the PCI controller and the motherboard nodes.

% TODO for thesis, show breakdown of bbus word. describe bus signals (CLK, DAT, STR) in more detail
Logically, the bus uses a command-response architecture. The PCI controller transmits a write word or a read request to a motherboard node, and then receives a response if necessary. Each 32-bit word contains 16 bits of data, and 16 bits of addressing and synchronization information. For a 4\,MHz main clock, this gives a data bandwidth of 1\,Mbit/s, which was chosen to match the line-of-sight radio transmitter used by balloon payloads. Writes and read requests are organized into a frame, which the PCI controller cycles through periodically---typically at around 100\,Hz, at an integer divisor of the data sample rate.

\begin{figure}
\centering
\includegraphics[width=0.95\linewidth]{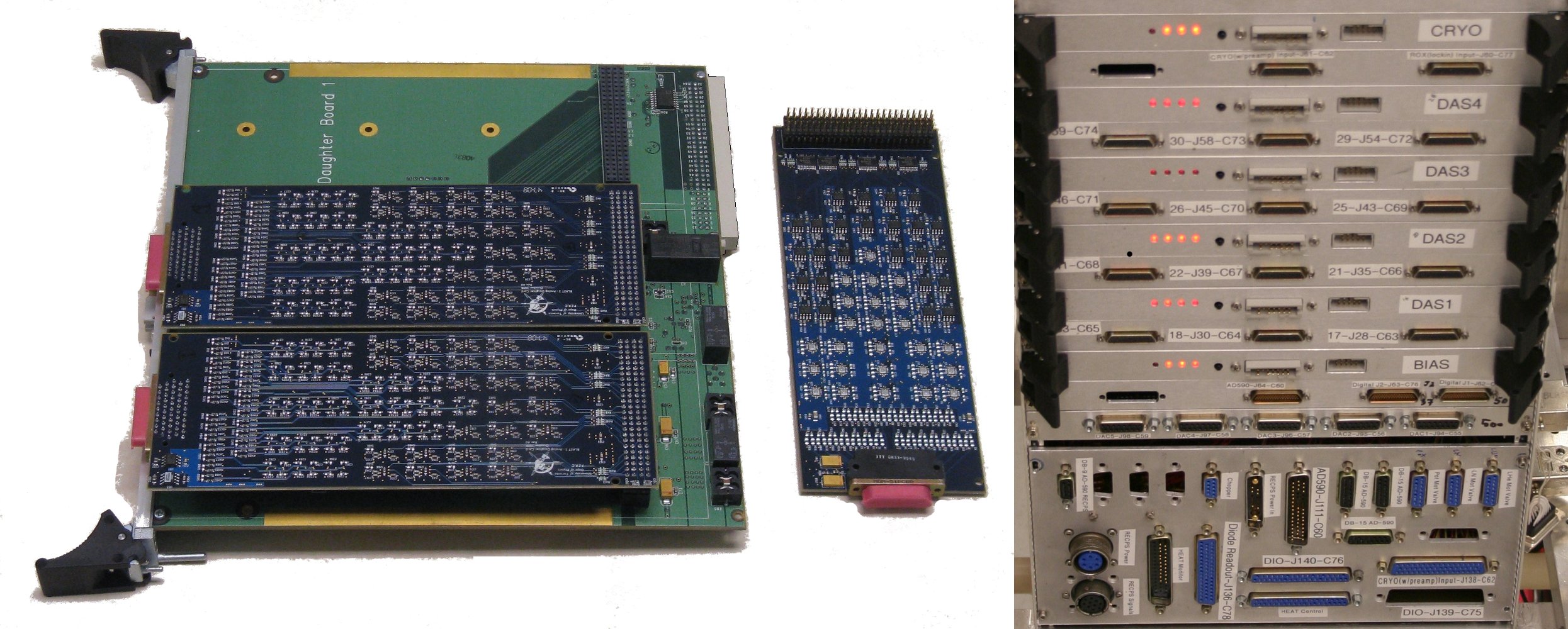}
\caption[BLASTBus motherboard, and DAS crate] 
{ \label{fig:blastbus_photo} 
Left: The underside of a BLASTbus motherboard, with one analog daughterboard removed. Right: The Data Acquisition System (DAS) crate on BLASTPol, with six BLASTbus motherboards, a DAC transmitter, and a breakout panel.}
\end{figure}

The BLASTbus is distributed to up to seven motherboards through the backplane of a 6\,U Eurocard crate. With careful impedance matching, the bus has been daisy-chained through two such crates. The crate also contains various power supplies for the motherboards. BLAST's data acquisition crate consumed 36\,W at 24\,VDC for six motherboards reading out 375 analog channels. Fig.~\ref{fig:blastbus_photo} shows images of a motherboard, three daughterboards, and a crate.

%\textbf{TODO} More power breakdown: ADCs (5mW each) about 2W total for DAS. Amps? DSP: 0.6W(mixed FP+SDRAM) to 0.9W(continuous FP). FPGA about 1.2W. Voltage regulators probably a bunch

\subsection{PCI Controller}
\label{sec:bbcpci}

The PCI controller board has a mainboard half for the logic and a daughterboard for the electrical interface. The mainboard contains an Altera Cyclone field-programmable gate array (FPGA), which implements the PCI layer, BLASTbus communications, and a NIOS II soft core for processing. The daughter board contains optoisolators and transceivers for two independent buses.

In addition to the BLASTbus, the PCI board generates a biphase encoded signal for the line-of-sight radio transmitter. An identical mainboard with slightly different firmware is used to receive the biphase signal on the ground.

\subsubsection{External synchronization}
\label{sec:sync}

The BLASTbus electronics are often\cite{rahlin2014,swetz2011,ogburn2010,staniszewski2012,kuo2013} used alongside the Multi-Channel Electronics (MCE) \cite{battistelli2008} for readout of Transition Edge Sensor (TES) bolometers\cite{irwin2005}. To synchronize the MCE and BLASTbus data, the PCI controller board has been adapted to interface to the MCE sync box in lieu of a second BLASTbus. This can either operate asynchronously, injecting synchronization data into the frame, or in a strongly synchronized mode where the BLASTbus clock is replaced by the MCE clock. In the strongly synchronized mode, the BLASTbus frame rate must be an integer divisor of the MCE frame rate.

\subsection{Motherboards}
\label{sec:motherboard}
The motherboards are the central hub of each BLASTbus slave node. In addition to the bus transceivers, they contain an Altera Cyclone II FPGA and an Analog Devices ADSP41369 digital signal processor (DSP). The DSP does the filtering and processing, while the FPGA presents a memory-mapped interface to the BLASTbus and the daughter boards. Each motherboard holds up to three daughter boards.

The motherboard cleans analog power for use on the daughter boards, using a combination of passive LC filtering and active voltage regulation. A watchdog circuit cycles all of the power supplies if the DSP fails to continuously toggle a watchdog reset line. Thick copper layers on the top and bottom conduct heat to the edges of the board. Due to a lack of convective cooling in the balloon environment, the heat conducting layers can have heat straps attached. In practice, radiation and air conduction provide sufficient cooling.

\subsection{Daughter Boards}
\label{sec:daughter_boards}

The BLASTbus system's input and output capabilities are determined by its daughter boards. In practice, only two types of daughter board have been needed: analog input and digital input/output. Analog outputs are generated by a secondary system connected to the digital daughter board.

Each analog board contains 25 low-power sigma-delta analog to digital converters (ADCs) with 24-bit resolution\footnote{Texas Instruments ADS1251}. These use the BLASTbus clock so that readout is fully synchronized (every 384 clock cycles, or 10.4\,kHz for a 4\,MHz clock). Typically, each ADC channel is paired with a preamp that has 50\,k$\Omega$ input impedance and accepts $\pm$10\,V. Alternatively, channels can be configured with no preamps, or with the capability to connect directly to thermistors or thermocouples. Each channel also has a low-pass RC filter at $\approx$60\,kHz to prevent aliasing of high frequency noise.

The digital daughter board contains six 8-bit optoisolated groups selectable as input or output. Input signals can be as fast as 15\,MHz, and a variety of firmware modules allow decoding of signals, such as digital gyroscopes or quadrature encoders. The isolated outputs can drive up to 50\,mA for switching relays, but are limited to speeds of about 30\,kHz.

The digital daughter board also contains an extra 8-bit group for commanding a digital to analog converter (DAC) output system. Each such system controls up to 32 DACs with $\pm$5\,V differential outputs and 16-bit resolution. These can update as fast as the analog input sample rate, and are also synchronized to the BLASTbus clock. Signals are distributed digitally as RS-485 to reduce noise, and so that the analog signal can be generated in a cleaner RF environment.

\subsection{Software and Firmware}
The BLASTbus system includes processing at three different levels: the control computer, the DSPs, and the FPGAs. The control computer is the most powerful and simple to program, but has access to data at only the $\approx$100\,Hz frame rate.

The bulk of the low-level data processing happens in the DSP. This operates in a real-time loop at the full $\approx$10\,kHz data sample rate. Processing includes anti-aliasing filters and lock-in amplification (Sec.~\ref{sec:ntd_readout}), and also application-specific fast control loops (eg. Sec.~\ref{sec:attitude} and Sec.~\ref{sec:housekeeping}).

The FPGA allows fast (80\,MHz) and extremely parallel processing of its signals. This is used for bit-level control of the ADCs, DACs, BLASTBus transceivers, and assorted digital sensors. Further processing could also take place in the FPGA, but in practice the DSP is simpler and adequate.

\section{NTD-G\lowercase{e} Bolometer Readout}
\label{sec:ntd_readout}

Neutron transmutation doped germanium (NTD-Ge) devices are sensitive low-temperature thermistors that are useful in millimetre and submillimetre bolometers\cite{turner2001}. Readout of BLAST's NTD-Ge bolometers was the primary performance driver for the BLASTbus system's analog readout.

\begin{figure}
\centering
\includegraphics[]{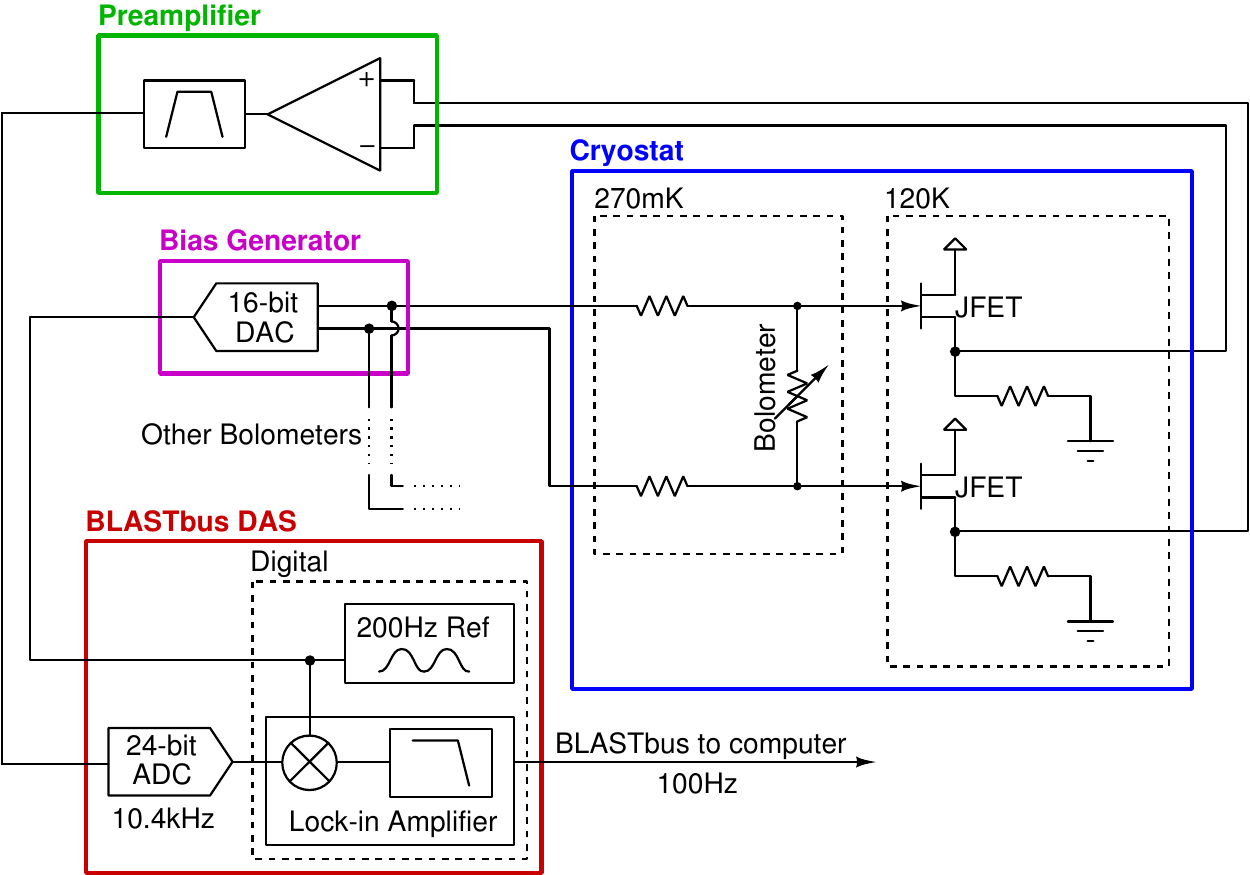}
\caption[NTD-Ge bolometer readout schematic] 
{ \label{fig:ntd_readout} 
Schematic view of the BLAST NTD-Ge bolometer readout. The BLASTbus data acquisition system generates a 200\,Hz sine-wave bias and digitally lock-in amplifies the signal from its ADCs. The cryogenic electronics include load resistors to quasi-current-bias the bolometer, and JFETs to reduce line impedance and microphonics. A preamplifier filters and conditions the signal for readout.}
\end{figure}

BLAST's bolometer readout system is sketched in Fig.~\ref{fig:ntd_readout}. Starting in the BLASTbus data acquisition system (DAS), a digital reference wave is generated.  This has a $\approx$200\,Hz frequency, which is exactly twice the frame rate, and faster than the bolometer time constant, but is still slow enough to be well-sampled by the $\approx$10\,kHz DAC update rate.

The bias generator is a simple DAC, which produces a differential sine wave. A passive voltage divider is then used so that a low-amplitude bias signal can be generated with higher amplitude and lower round-off error at the DAC. The voltage divider has low resistance to minimize Johnson noise, and a small current amplifier is required for the DAC to drive it. The bias signal is shared by all bolometers in each of BLAST's three detector arrays, but the bias amplitude is tuned differently between arrays. A fourth bias signal is used for the cryogenic thermistors (see Sec.~\ref{sec:housekeeping}).

The voltage bias produced by the DAC is converted to a nearly constant current bias by load resistors (typically $\approx$8\,M$\Omega$), which are larger than the bolometers' nominal resistance ($\approx$1.5\,M$\Omega$), and much larger than changes in bolometer resistance during operation. The final cryogenic element is a pair of U401 JFET followers to lower the impedance and reduce microphonics due to the capacitance of vibrating cables. These are placed as close to the bolometers as possible, with the cables in between tightly constrained. The thermally isolated JFET enclosure is self-heated to 140\,K, which locally minimizes noise and exceeds their freeze-out temperature of $\approx$100\,K.

The JFETs' signals then pass through the ambient temperature preamplifier, which is connected to the cryostat by a shielded cable bundle. The preamplifier filters the data using a biquad band-pass with 85\,Hz bandwidth around the 200\,Hz bias frequency. To save power, the preamplifier conditions the signal for direct acquisition by the BLASTbus ADCs, without an extra amplifier on the daughterboard.

\begin{figure}
\centering
\includegraphics[]{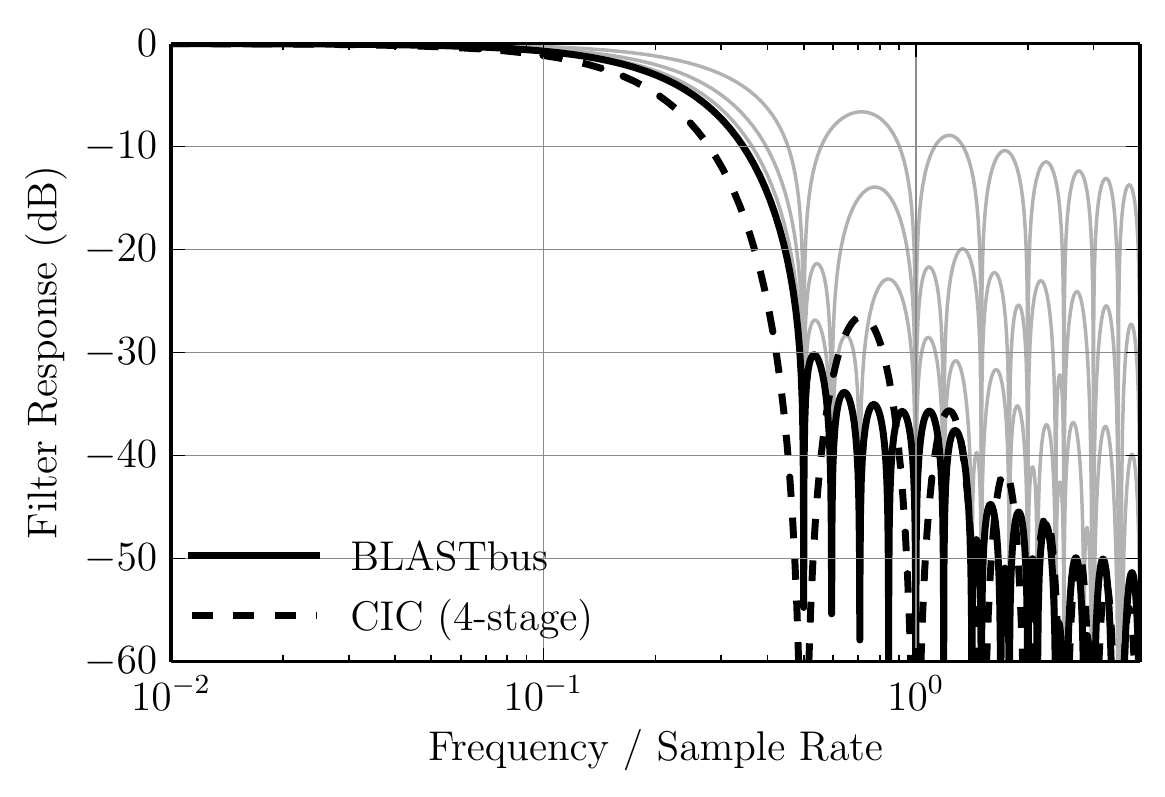}
\caption[BLASTbus digital filter response] 
{ \label{fig:filter} 
Frequency response of the BLASTbus digital anti-aliasing filter. For comparison, a 4-stage CIC filter is shown, with more passband droop and increased sidelobes. The cumulative response at intermediate filter stages is also shown in light grey}
\end{figure}

Finally, the digital signal is mixed with the reference wave and anti-alias filtered for acquisition. The digital filter, shown in Fig.~\ref{fig:filter}, leaves just the DC mixer output, which is reported over the BLASTbus and stored to disk. Because the bias is synced to the frame rate, the filter has a null at the bias frequency and the 2f mixer output is strongly removed. The digital filter itself is a 4-stage boxcar filter with first-nulls spaced logarithmically between the Nyquist frequency and sample rate. This results in a flatter passband and lower sidelobes than a 4-stage CIC (cascaded integrator-comb) filter\cite{hogenauer1981}.

Like a CIC filter, this filter uses just addition and subtraction at each stage, though it requires more memory as data cannot all be stored at the decimated rate. For modest decimation of 104 samples, this is not a problem. For the BLASTPol DAS, with 375 channels, there is not enough bandwidth to acquire both the sine and cosine components of the signal. Instead, just the cosine component is acquired and a phase is set to maximize its amplitude, and minimize the sine component.

\subsubsection{Performance}

\begin{figure}
\centering
\includegraphics[]{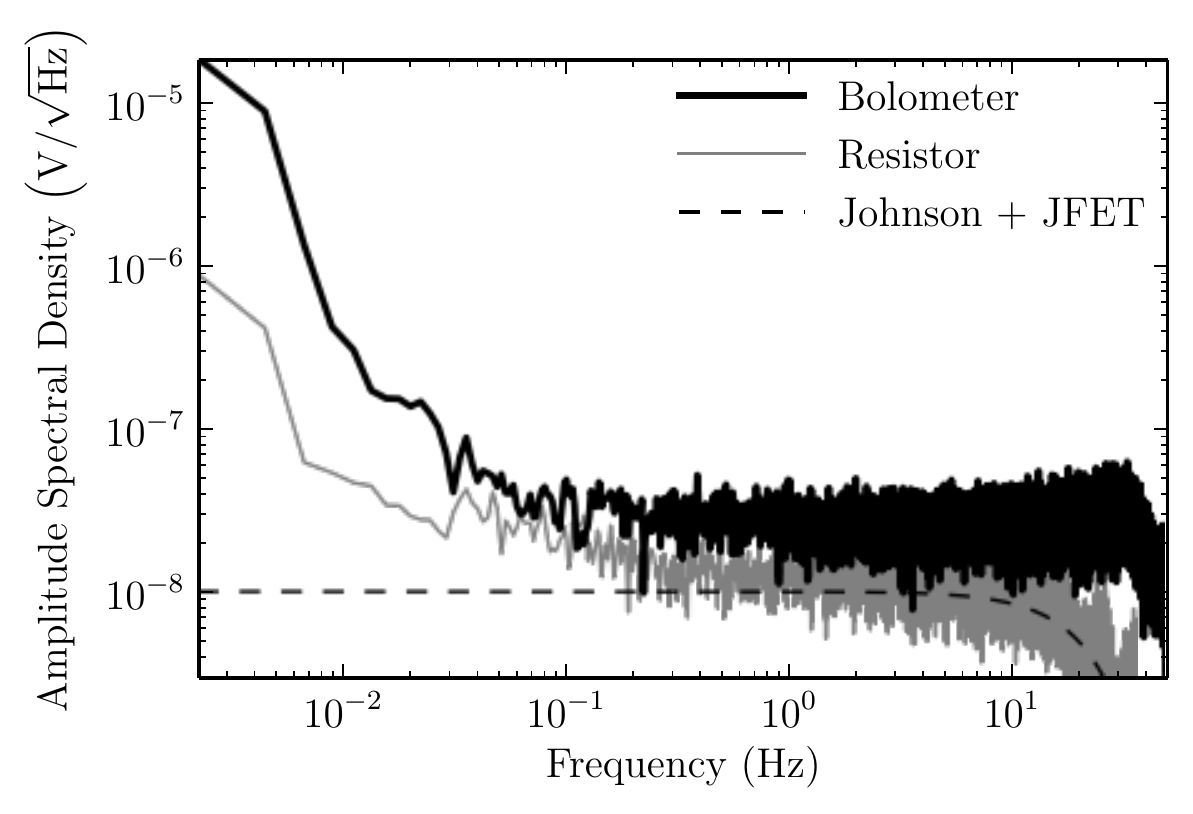}
\caption[BLAST bolometer noise] 
{ \label{fig:bolo_spectrum}
Example bolometer noise spectrum from BLAST's 2012 flight. Also shown is the spectrum of a 4\,M$\Omega$ fixed resistor, used to measure electronic noise. The resistor's white noise level is close to its Johnson noise, with smaller contributions from the JFETs and warm amplifier. The higher noise of the bolometer is mostly due to photon shot noise. These spectra were taken while the telescope was not scanning, so contain little signal. The bolometer data have been despiked and have had the anti-aliasing filter deconvolved.}
\end{figure}

The digital lock-in achieves about 20 bits of dynamic range when looped back into itself, without the bolometers, JFETs, or preamplifier. Fig.~\ref{fig:bolo_spectrum} shows full-system noise spectra from the 2012 flight of BLASTPol. Reading a fixed cryogenic resistor, Johnson noise is the largest contributor, with the remainder due to the JFETs and warm amplifier. Most importantly, the bolometer noise is dominated by photon shot noise, which exceeds all sources of electrical and readout noise.

\section{ATTITUDE CONTROL}
\label{sec:attitude}

Attitude control involves reading a variety of sensors, commanding motors, and then tying both tasks together into a control loop. With its flexible inputs, programmability, and real-time processing, the BLASTbus system is well suited to these tasks. Attitude control is a major application of the BLASTbus electronics, but because the topic is covered by two other papers in these proceedings\cite{gandilo2014,shariff2014}, only a summary is presented here.

Many types of sensor are used to measure attitude, both during the flight and in post-flight processing. Most have their signals acquired directly by the BLASTbus, via analog or digital daughterboards. For the star trackers, the BLASTbus is used only to synchronize camera exposures with other data. The primary sensor for control purposes is a set of fiber-optic gyroscopes, measuring rotation rate around three axes. The other sensors serve to provide absolute position information with varying accuracies, sample rates, and reliabilities. To minimize delay, single-stage filtered gyroscope streams are used in the real-time loop, while fully-filtered data are stored for post-processing. More details can be found in Ref.~\citenum{gandilo2014}.

BLAST and \textsc{Spider} use three motors to control attitude: an elevation drive, a reaction wheel for fine azimuth, and an actuated pivot connecting to the balloon for coarse azimuth. The BLASTbus attitude control system commands the motor controllers via its DAC outputs. High-level logic runs on the master control computer, sending set-points and data to the BLASTbus board at the 100\,Hz frame rate. In between these updates, the DSP servos the motors using its faster data---mainly the gyroscopes. Details of the scan strategy and the control loops are in Ref.~\citenum{shariff2014}.

Polarimeters like \textsc{Spider} and BLASTPol feature another attitude-like quantity: the orientation of half-wave plates, which modulate the input polarization signal. BLASTPol's half-wave plate\cite{moncelsi2014} is surrounded by a rotating potentiometer, which is DC biased and read out as a simple analog signal. Spider's half-wave plates use contact-free optical encoders\cite{bryan_2014_thesis}. The LEDs in this system are square-wave biased at 1\,kHz using BLASTbus digital outputs, and the reflected photodiode signals are digitally demodulated much like in the NTD-Ge bolometer readout (Sec.~\ref{sec:ntd_readout})

\section{CRYOGENIC HOUSEKEEPING}
\label{sec:housekeeping}

While \textsc{Spider} does not use NTD-Ge bolometers, it does have cryogenic thermistors---including NTD-Ge devices at sub-Kelvin temperatures. To read these, wide dynamic range is important, as the resistance can vary by several orders of magnitude over the operating temperature range. At higher temperatures, simpler diode thermometers can be used. To manipulate the cryogenic environment, it is also necessary to control a variety of heaters.

With six independent cryogenic telescopes, and balloon-payload power constraints, \textsc{Spider} needed a simple and effective means to perform these tasks. A unified board was made to handle the needs of each individual telescope. This uses the same form factor, crate, and power cleaning as the BLASTbus motherboards, and provides cryogenic housekeeping with a minimum of extra capabilities on top of what the BLASTbus system supplies already. Each board has circuits for 8 thermistors, 18 diode thermometers, 8 digital heaters, and 2 analog heaters.

\begin{figure}
\centering
\includegraphics[]{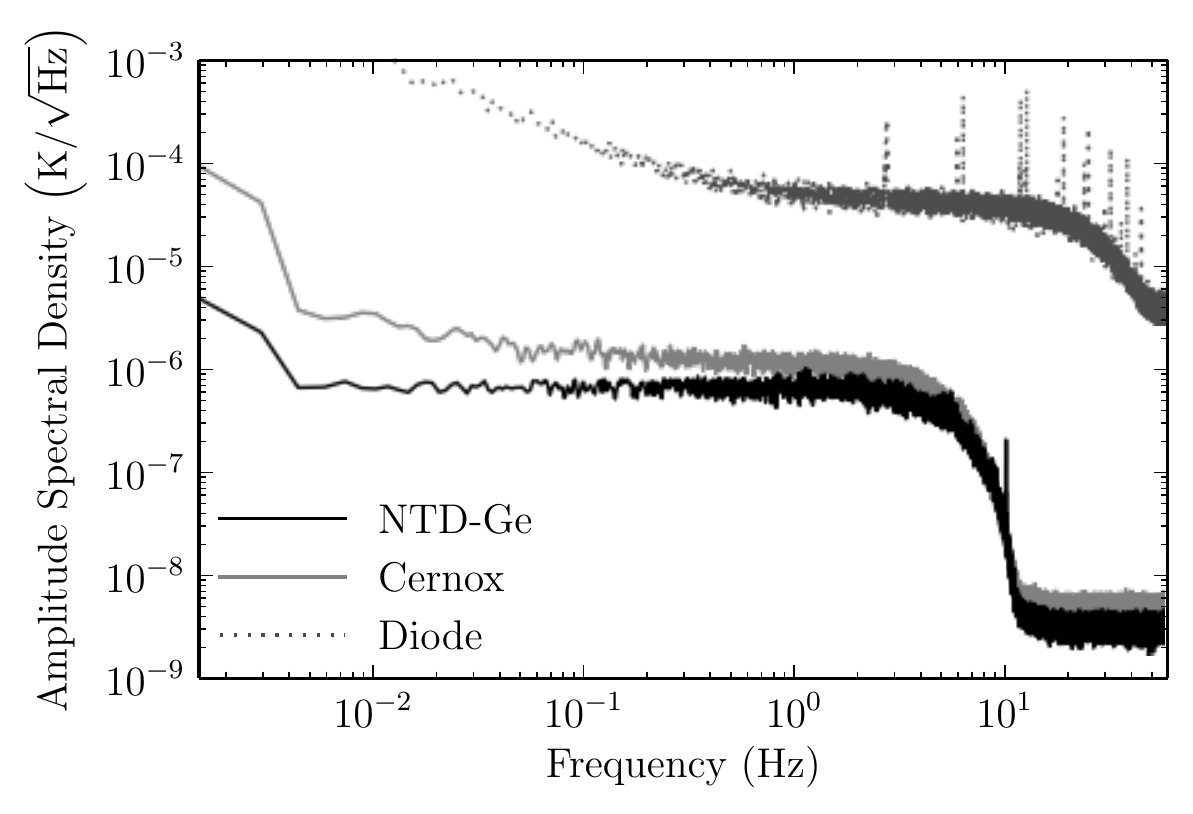}
\caption[Noise levels of cryogenic thermometers] 
{ \label{fig:hk_noise}
Noise spectra of \textsc{Spider} cryogenic thermometers, in temperature units. While the NTD-Ge and Cernox thermistors perform similarly, the NTD-Ge device has only 7\% as much bias current, to prevent heating of the detector tile. The thermistors are biased at 10\,Hz, so their filters cut off well below the Nyquist frequency.}
\end{figure}

The thermistor readout is similar to the NTD-Ge bolometer readout scheme, though somewhat simplified as it does not need to aggressively lower noise. The main simplification is to not include cold JFETs and their isolated cavity. Instead, warm JFETs are used, integrated into the input of the amplifier. The bias frequency is made commandable to help find a clean frequency that doesn't contaminate other systems. Typically, the bias frequency is in the range of 10--100\,Hz. To accommodate changing frequencies, the bandpass filter used in bolometer readout is replaced by a 5\,Hz highpass filter and 500\,Hz lowpass filter. With only 50 thermistor channels, full quadrature lock-in is possible, eliminating the need to tune the phase. Noise spectra, calibrated into temperature units, are shown in Fig.~\ref{fig:hk_noise}.

Diode thermometers are used where the higher precision of thermistors is not necessary, and at temperatures above about 1\,K\@. They are biased by a temperature-corrected 10\,$\mu$A DC current source and their temperature-dependent voltage is amplified slightly before acquisition by the BLASTbus system. The diode circuit uses a non-differential 2-wire bias/readout scheme. An example of the diode noise level is also shown in Fig.~\ref{fig:hk_noise}.

In common cases, like recycling the helium-3 adsorption fridge, a coarse on/off heater is desired, and these are provided with digital control using a solid-state relay. The relay output voltage and heater resistance can be chosen to set the power level. To coarsely adjust the power, these can also be driven by a pulse width modulated signal. For finely adjustable heating, filtered pass-throughs are available to drive heaters directly with a DAC. BICEP2 and the Keck Array use both types of heaters, passive thermal filters, and feedback from NTD-Ge thermistors to stabilize their focal plane temperatures\cite{bicep2_2014_instrument}. BICEP2 achieves a stability of 0.4\,nK at $\ell$ of 100\footnote{This is a measure of data contamination due to uncorrected temperature fluctuations, averaged over detectors and time. The fluctuations are smaller than the NTD-Ge noise and difficult to measure directly}.

\section{CONCLUSION}
With a simple programmable architecture and many parallel analog and digital channels, the BLASTbus electronics can simultaneously accomplish many precision readout and control tasks. They are a low-power all-in-one solution used throughout BLASTPol, but are also effectively employed solely for gondola attitude control or cryogenic housekeeping. So far, only general-purpose daughter boards have been used, sometimes with auxilliary circuits. Where appropriate, specialized daughter boards could be created that offer fully integrated solutions for a variety of applications.

%%%%%%%%%%%%%%%%%%%%%%%%%%%%%%%%%%%%%%%%%%%%%%%%%%%%%%%%%%%%%
\acknowledgments     %>>>> equivalent to \section*{ACKNOWLEDGMENTS}
The BLASTbus electronics have benefited from the expertise and efforts of specialists at the University of Toronto's Physics Electronics Resource Center.

The \textsc{Spider} collaboration gratefully acknowledges the support of NASA (award numbers NNX07AL64G, NNX12AE95G), the Lucille and David Packard Foundation, and the Gordon and Betty Moore Foundation. The BLAST collaboration acknowledges the support of NASA through grant numbers NNX13AE50G S03 and NNX09AB98G, the Leverhulme Trust through the Research Project Grant F/00 407/BN. We further acknowledge the support of the Natural Sciences and Engineering Research Council (NSERC), the Canadian Space Agency (CSA), the Canada Foundation for Innovation, the Ontario Innovation Trust, the Puerto Rico Space Grant Consortium, the Fondo Institucional para la Investigacion of the University of Puerto Rico, the Rhode Island Space Grant Consortium. F.~Poidevin thanks the Spanish Ministry of Economy and Competitiveness (MINECO) under the Consolider-Ingenio project CSD2010-00064 (EPI: Exploring the Physics of Inflation). W.~C.~Jones acknowledges the support of the Alfred P. Sloan Foundation. A.~S.~Rahlin is partially supported through NASAs NESSF Program (12-ASTRO12R-004). J.~D. Soler acknowledges the support of the European Research Council under the European Union's Seventh Framework Programme FP7/2007-2013/ERC grant agreement number 267934. We thank the JPL Research and Technology Development Fund for advancing detector focal plane technology. Logistical support for this project in Antarctica is provided by the U.S. National Science Foundation through the U.S. Antarctic Program. We would also like to thank the Columbia Scientific Balloon Facility (CSBF) staff for their continued outstanding work. 

%%%%%%%%%%%%%%%%%%%%%%%%%%%%%%%%%%%%%%%%%%%%%%%%%%%%%%%%%%%%%
%%%%% References %%%%%

\bibliography{blastbus}   %>>>> bibliography data in report.bib
\bibliographystyle{spiebib}   %>>>> makes bibtex use spiebib.bst

\end{document}